\documentclass[a4paper,english,journal,10pt,twocolumn]{IEEEtran}
\usepackage[T1]{fontenc}
\usepackage[latin9]{inputenc}
\usepackage{color}
\usepackage{textcomp}
\usepackage{amsthm}
\usepackage{amsmath}
\usepackage{amssymb}
\usepackage{graphicx}
\DeclareGraphicsExtensions{.eps,.pdf}

\makeatletter

\providecommand{\tabularnewline}{\\}

\theoremstyle{plain}
\newtheorem{thm}{\protect\theoremname}
\theoremstyle{definition}
\newtheorem{defn}[thm]{\protect\definitionname}

\usepackage[font={small}]{caption}\usepackage{changebar}\usepackage[noadjust]{cite}\usepackage{array}\usepackage[ruled,lined,linesnumbered,longend]{algorithm2e}
\usepackage{url}
\providecommand{\definitionname}{Definition}
\providecommand{\theoremname}{Theorem}

\providecommand{\definitionname}{Definition}
\providecommand{\theoremname}{Theorem}

\providecommand{\definitionname}{Definition}
\providecommand{\theoremname}{Theorem}

\@ifundefined{showcaptionsetup}{}{%
 \PassOptionsToPackage{caption=false}{subfig}}
\usepackage{subfig}
\makeatother

\providecommand{\definitionname}{Definition}
\providecommand{\theoremname}{Theorem}

\begin{document}

\title{Joint Transmit and Receive Beamforming for Multi-Relay MIMO-OFDMA Cellular Networks}

\author{Kent Tsz Kan Cheung, Shaoshi Yang and Lajos Hanzo\\
{\small{} School of ECS, University of Southampton, SO17 1BJ, United
Kingdom.}\\
{\small{} Email: \{ktkc106,sy7g09,lh\}@ecs.soton.ac.uk, wireless.ecs.soton.ac.uk}\vspace{-10.5mm}
}

\markboth{Accepted by IEEE ICC 2016, 23-27 May 2016, Kuala Lumpur, Malaysia}%
{Shell \MakeLowercase{\textit{et al.}}: Bare Demo of IEEEtran.cls
for Journals}

\maketitle

\begin{abstract}
A novel transmission protocol is conceived for a multi-user,
multi-relay, multiple-input--multiple-output orthogonal frequency-division
multiple-access~(MIMO-OFDMA) cellular network based on joint transmit and
receive beamforming. More specifically, the network's MIMO channels are mathematically decomposed
into several effective multiple-input--single-output~(MISO) channels,
which are spatially multiplexed for transmission.
For the sake of improving the attainable capacity, these MISO channels are
grouped using a pair of novel grouping algorithms, which are then evaluated in terms of their performance
versus complexity trade-off\footnote{This paper concisely focuses on the transmission protocol proposed in our previous work~\cite{Cheung2014}. For more details, please refer to~\cite{Cheung2014}.}.\end{abstract}
\vspace{-5mm}
\section{Introduction\label{sec:Intro}}

Recent wireless mobile broadband standards optionally
employ relay nodes~(RNs) and multiple-input--multiple-output orthogonal
frequency-division multiple-access~(MIMO-OFDMA) systems~\cite{Salem2010,Hanzo2010}
for supporting the ever-growing wireless capacity demands. These systems
benefit from a capacity gain increasing roughly linearly both with
the number of available OFDMA subcarriers~(each having the same bandwidth)
as well as with the minimum of the number of transmit antennas~(TAs)
and receive antennas~(RAs). However, given the additional
resources, the issue arises as to how best to allocate them for maximizing
the system's capacity. \emph{In light of these discussions, we propose
a novel joint transmit and receive beamforming~(BF) protocol for coordinating
the downlink~(DL) transmissions in a sophisticated multi-relay aided MIMO-OFDMA cellular network.}

It is widely acknowledged that under the idealized simplifying condition
of having perfect channel state information~(CSI) at the transmitter,
the DL or broadcast channel~(BC) capacity~\cite{Caire2003,Vishwanath2003}
may be approached with the aid of dirty paper coding~(DPC)~\cite{Costa1983}.
However, the practical implementation of DPC
is hampered by its excessive algorithmic complexity upon increasing
the number of users. On the other hand, BF is an attractive suboptimal
strategy for allowing multiple users to share the BC while resulting
in reduced multi-user interference~(MUI). A low-complexity transmit-BF
technique is the zero-forcing based BF~(ZFBF), which can asymptotically
achieve the BC capacity as the number of users tends to infinity~\cite{Yoo2006}.
Furthermore, ZFBF may be readily applied to a system with multiple-antenna
receivers through the use of the singular value decomposition~(SVD).
As a result, the associated MIMO channels may be
mathematically decomposed into several \emph{effective} multiple-input--single-output~(MISO)
channels, which are termed spatial multiplexing components~(SMCs) in this work.

In~\cite{UlHassan2009}, these SMCs are
specifically grouped so that the optimal grouping as well as the optimal
allocation of the power may be found on each subcarrier
block using convex optimization. In contrast to the
channel-diagonalization methods of~\cite{Raleigh1998,Wong2003,Ho2009},
the ZFBF approach does not enforce any specific relationship between
the total numbers of TAs and RAs. Therefore, ZFBF is more suitable
for practical systems, since the number of TAs at the BS is typically
much lower than the total number of RAs of all the active user equipments~(UEs).
Compared to the random beamforming methods, such as that
of~\cite{Sharif2005}, ZFBF is capable of completely
avoiding the interference, thus improving the system's attainable capacity.

Due to its desirable performance versus complexity
trade-off, in this paper we employ
ZFBF in the context of multi-relay aided MIMO-OFDMA systems, where
the direct link between the base station~(BS) and the UE may be exploited
in conjunction with the relaying link for further improving the system's
performance.

In this paper, we propose a novel transmission protocol for a generalized multi-user
multi-relay aided MIMO-OFDMA cellular system, which supports simultaneous direct
and relayed transmissions without imposing interference on the receivers. This is
accomplished by mathematically decomposing the network's channel matrices
for ensuring that the beneficial links may be grouped for simultaneous transmission.
The system model in~\cite{Yoo2006,UlHassan2009,Brante2013,Zappone2014} is improved, since these contributions
did not consider exploiting relaying for improving the system's performance, or only considered
single-relay, single-user scenarios. Furthermore, we imposed no constraint on the relationship between
the number of TAs and RAs in the system, which was assumed in~\cite{Raleigh1998,Wong2003,Ho2009}.

Furthermore, we conceive a pair of novel algorithms for grouping the SMCs transmissions.
These challenging issues of two-phase communication in the presence of multiple transmitters as well as simultaneous direct and
relayed transmissions are resolved by the proposed grouping algorithms.
The first grouping algorithm is optimal in the sense that
it is based on exhaustive search over all the SMC
groupings that satisfy the semi-orthogonality criterion, while the
second algorithm constitutes a lower-complexity alternative. In terms of its basic principle,
the lower-complexity method is reminiscent of~\cite{Yoo2006,UlHassan2009},
but it has been appropriately adapted for the multi-relay cellular
network considered.
\vspace{-5mm}
\section{System model\label{sec:SysModel}}

\begin{figure}
\centering
\includegraphics[scale=0.54]{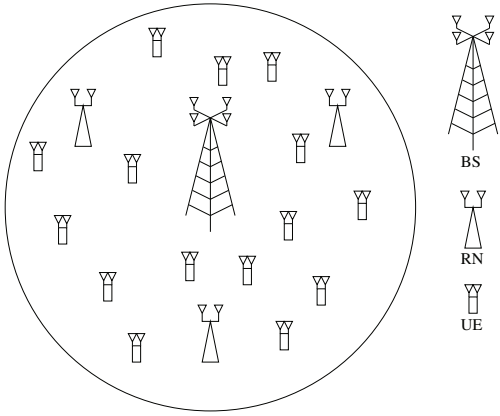} 
\caption{An example of a multi-relay MIMO-OFDMA cellular network, containing a BS at the cell-center, 3 RNs and 15 UEs.}
\label{fig:cellular} 
\vspace{-7mm}
\end{figure}

We focus our attention on the DL of a multi-relay MIMO-OFDMA cellular
network, as shown in Fig.~\ref{fig:cellular}. The BS, $M$
RNs and $K$ UEs are each equipped with $N_{B}$, $N_{R}$ and $N_{U}$
antennas, respectively. The cellular system has access to $N$ subcarrier
blocks, each encompassing $W$ Hertz of wireless bandwidth. The subcarrier
blocks considered here are similar to the resource blocks in the LTE-nomenclature~\cite{3GPP_PL}.
The BS is located at the cell-center, while the RNs are each located
at a fixed distance from the BS and are evenly spaced around it. On the other hand, the UEs are uniformly distributed
in the cell. The BS coordinates and synchronizes its own transmissions
with that of the RNs, which employ the decode-and-forward~(DF)~\cite{Laneman2004} transmission protocol and
thus avoids the problem of noise amplification.

For the subcarrier block $n\in\left\{ 1,\cdots,N\right\} $, let us
define the complex-valued wireless channel matrices between the BS
and UE $k\in\left\{ 1,\cdots,K\right\} $, between the BS and RN $m\in\left\{ 1,\cdots,M\right\} $,
and between RN $m$ and UE $k$ as $\mathbf{H}_{n,k}^{BU}\in\mathbb{C}^{N_{U}\times N_{B}}$,
$\mathbf{H}_{n,m}^{BR}\in\mathbb{C}^{N_{R}\times N_{B}}$ and $\mathbf{H}_{n,m,k}^{RU}\in\mathbb{C}^{N_{U}\times N_{R}}$,
respectively. These complex-valued channel matrices account for both
the frequency-flat Rayleigh fading and the path-loss between the corresponding
transceivers. The coherence bandwidth of each wireless link is assumed
to be sufficiently high, so that each individual subcarrier block
experiences frequency flat fading, although the level of fading may
vary from one subcarrier block to another
in each transmission period. Additionally, the transceivers are stationary
or moving slowly enough so that the level of fading may
be assumed to be fixed for the duration of a scheduled
transmission period. Furthermore, the RAs are spaced sufficiently
far apart, so that each TA/RA pair experiences independent and identically
distributed~(i.i.d.) fading. Since these channels are slowly varying,
the system is capable of exploiting the benefits of channel
reciprocity associated with time-division duplexing~(TDD) as well
as dedicated low-rate error-free feedback channels,
so that the CSI becomes available at each BS- and RN-transmitter as well as
at each possible RN- and UE-receiver. It is assumed that the BS performs
network-wide scheduling and that these channel matrices have full row rank, which
may be achieved with a high probability for typical DL wireless channel
matrices.

Furthermore, each receiver suffers from additive white Gaussian noise~(AWGN)
having a power spectral density of $N_{0}$. The maximum instantaneous
transmission power available to the BS and to each RN due to regulatory
and health-constraints is $P_{max}^{B}$ and $P_{max}^{R}$, respectively.
Since OFDMA modulation constitutes a linear operation, we focus our
attention on a single subcarrier block and as usual, we employ the
commonly-used equivalent baseband signal model
\vspace{-3mm}
\section{Transmission protocol design\label{sec:TxProt}}

The system can simultaneously use two transmission modes to convey
information to the UEs, namely the BS-to-UE mode, and the relaying-based
BS-to-RN and RN-to-UE mode. Note that although in classic OFDMA each
data stream is orthogonal in frequency, for the sake of further improving
the system's attainable performance, our system employs
spatial multiplexing in conjunction with ZFBF \emph{so that multiple
data streams may be served using the same subcarrier block, without
suffering from interference}. Additionally, since the relaying-based
transmission can be split into two phases, the design philosophy of
the BF matrices in each phase are described separately, although for
simplicity we have assumed that the respective channel matrices remain
unchanged in both phases. Firstly,
the definition of the semi-orthogonality criterion is given as follows~\cite{Yoo2006}. 
\begin{defn}
A pair of MISO channels, represented by the complex-valued column
vectors $\mathbf{v}_{1}$ and $\mathbf{v}_{2}$, are said to be semi-orthogonal
to each other with parameter $\alpha\in\left[0,1\right]$, when\footnote{In this paper, $\Re\left(x\right)$ denotes the real component of $x$.}
\begin{equation}
\frac{\left|\Re\left(\mathbf{v}_{1}^{\mathrm{H}}\mathbf{v}_{2}\right)\right|}{\|\mathbf{v}_{1}\|\|\mathbf{v}_{2}\|}\leq\alpha.\label{eq:semi-ortho}
\end{equation}
To be more specific, a measure of the grade of orthogonality between
$\mathbf{v}_{1}$ and $\mathbf{v}_{2}$ is given by the left-hand
side of inequality~(\ref{eq:semi-ortho}), which ranges from $0$
for orthogonal vectors to $1$ for linearly dependent vectors. 
\end{defn}
The authors of~\cite{Yoo2006} demonstrated that employing the ZFBF
strategy for MISO channels that satisfy $\alpha\to0$, while the number
of users obeys $K\to\infty$, asymptotically achieves the DPC capacity,
and it is therefore optimal for the BC channel. Similar principles
are followed in this paper.
\vspace{-3mm}
\subsection{BF design for the first transmission phase\label{sub:BFT1}}

In the first transmission phase, only the BS is transmitting, while
both the RNs and the UEs act as receivers. This is similar to the
classic DL multi-user MIMO model. As described above, our aim is 1)
to design a ZFBF matrix for the BS to avoid interference between data
streams, and 2) to design receive BF matrices for
the UEs and RNs so that the resultant \emph{effective DL channel matrices}
contain as many semi-orthogonal
rows as possible that satisfy~(\ref{eq:semi-ortho})
for a given $\alpha$. Ideally, a joint computation of the receive BF
matrices is performed for satisfying the latter condition. However, this is generally
impossible due to the geographically distributed nature of the UEs and RNs. Instead, we opt for
a compromise by employing the SVD, so that at least each individual effective DL channel
matrix contains orthogonal rows.

Bearing this in mind, the channel matrices of all DL transmissions
originating from the BS are decomposed at the BS,
UEs and RNs using the SVD as $\mathbf{H}_{n,k}^{BU}=\mathbf{U}_{n,k}^{BU}\mathbf{S}_{n,k}^{BU}\left(\mathbf{V}_{n,k}^{BU}\right)^{\mathrm{H}}$
and $\mathbf{H}_{n,m}^{BR}=\mathbf{U}_{n,m}^{BR}\mathbf{S}_{n,m}^{BR}\left(\mathbf{V}_{n,m}^{BR}\right)^{\mathrm{H}}$,
respectively. Thus, the receive-BF matrices for UE $k$ and RN $m$
are given by $\mathbf{R}_{n,k}^{BU,T_{1}}=\left(\mathbf{U}_{n,k}^{BU}\right)^{\mathrm{H}}$
and $\mathbf{R}_{n,m}^{BR,T_{1}}=\left(\mathbf{U}_{n,m}^{BR}\right)^{\mathrm{H}}$,
and the effective DL channel matrices are then given%
\footnote{Note that $T_{1}$ is used for indicating the first transmission phase,
and underline is used to denote the effective DL
channel matrices.%
} by $\underline{\mathbf{H}}_{n,k}^{BU,T_{1}}=\mathbf{R}_{n,k}^{BU,T_{1}}\mathbf{H}_{n,k}^{BU}=\mathbf{S}_{n,k}^{BU}\left(\mathbf{V}_{n,k}^{BU}\right)^{\mathrm{H}}$
and $\underline{\mathbf{H}}_{n,m}^{BR,T_{1}}=\mathbf{R}_{n,m}^{BR,T_{1}}\mathbf{H}_{n,m}^{BR}=\mathbf{S}_{n,m}^{BR}\left(\mathbf{V}_{n,m}^{BR}\right)^{\mathrm{H}}$,
respectively. Since $\mathbf{V}_{n,k}^{BU}$ and $\mathbf{V}_{n,m}^{BR}$
are both unitary, while $\mathbf{S}_{n,k}^{BU}$ and $\mathbf{S}_{n,m}^{BR}$
are both real and diagonal, these effective DL
channel matrices respectively consist of $\min\left(N_{B},N_{U}\right)$
and $\min\left(N_{B},N_{R}\right)$ orthogonal non-zero rows with norms equal to their corresponding singular values. We refer
to these non-zero orthogonal rows as the SMCs of their associated
MIMO channel matrix. The $K$ BS-to-UE MIMO channel matrices and $M$ BS-to-RN channel
matrices generate a total of $\left[K\cdot\min\left(N_{B},N_{U}\right)+M\cdot\min\left(N_{B},N_{R}\right)\right]$
SMCs. Since these SMCs are generated from independent MIMO channel
matrices associated with geographically distributed UEs and RNs, they
are not all guaranteed to be orthogonal to each other. Furthermore,
since each UE or RN has multiple antennas and $N_{B}$ might not be
sufficiently large to simultaneously support all UEs and RNs, we have
to determine which specific SMCs should be selected. As a result, for
each two-phase transmission period, we opt for selecting a SMC group
accounting for both phases from the set of available SMC groups. The generation
of SMC groups is accomplished by the SMC grouping algorithms to be described in Section~\ref{sec:SGAlgor}.

To elaborate a little further, a set of SMC groups, $\mathcal{G}_{n}$,
which is associated with subcarrier block $n$,
may be obtained using one of the grouping algorithms presented in
Section~\ref{sec:SGAlgor}. The BS selects a single group, $j\in\mathcal{G}_{n}$,
containing~(but not limited to%
\footnote{The SMC group selection, as a part of the scheduling operation, is
carried out at the BS before initiating the first transmission phase.
Hence, the selected SMC group will also contain $Q_{j}^{T_{2}}$ SMCs
selected by the BS for the second transmission phase, as detailed
in Section~\ref{sub:BFT2}.
}) $Q_{j}^{T_{1}}$ SMCs out of the $\left[K\cdot\min\left(N_{B},N_{U}\right)+M\cdot\min\left(N_{B},N_{R}\right)\right]$
available SMCs to be supported by using ZFBF. Thus, we have $Q_{j}^{T_{1}}\leq\min\left[N_{B},K\cdot\min\left(N_{B},N_{U}\right)+M\cdot\min\left(N_{B},N_{R}\right)\right]$
and a multiplexing gain of $Q_{j}^{T_{1}}$ is achieved.
Let us denote the refined effective DL channel matrix
with rows being the $Q_{j}^{T_{1}}$ selected SMCs as $\underline{\mathbf{H}}_{n,j}^{T_{1}}\in\mathbb{C}^{Q_{j}^{T_{1}}\times N_{B}}$.
The ZFBF transmit matrix applied at the BS to subcarrier block $n$
is then given by the following right inverse $\mathbf{T}_{n,j}^{T_{1}}=\left(\underline{\mathbf{H}}_{n,j}^{T_{1}}\right)^{\mathrm{H}}\cdot\left[\underline{\mathbf{H}}_{n,j}^{T_{1}}\left(\underline{\mathbf{H}}_{n,j}^{T_{1}}\right)^{\mathrm{H}}\right]^{-1}$.
Since $\underline{\mathbf{H}}_{n,j}^{T_{1}}\mathbf{T}_{n,j}^{T_{1}}=\mathbf{I}_{N_{B}}$,
the potential interference between the $Q_{j}^{T_{1}}$
selected SMCs is completely avoided. Furthermore,
the columns of $\mathbf{T}_{n,j}^{T_{1}}$ are normalized by multiplying
the diagonal matrix $\mathbf{W}_{n,j}^{T_{1}}$ on the right-hand
side of $\mathbf{T}_{n,j}^{T_{1}}$ to ensure that each SMC transmission
is initially set to unit power.

Then, $\mathbf{T}_{n,j}^{T_{1}}\mathbf{W}_{n,j}^{T_{1}}$ is used
as the DL transmit-BF matrix for the BS in the first
phase. Thus, the effective channel-to-noise ratios~(CNRs) in the
first transmission phase can be written as $G_{n,j,e_{1}}^{BU,T_{1}}=\left|w_{n,j,e_{1}}^{BU,T_{1}}\right|^{2}/\Delta\gamma N_{0}W$
and $G_{n,j,e}^{BR,T_{1}}=\left|w_{n,j,e}^{BR,T_{1}}\right|^{2}/\Delta\gamma N_{0}W$,
respectively, where $w_{n,j,e_{1}}^{BU,T_{1}}$ and $w_{n,j,e}^{BR,T_{1}}$
are the diagonal elements in $\mathbf{W}_{n,j}^{T_{1}}$, $\Delta\gamma$ is the
signal-to-noise ratio~(SNR) gap, and noise power received on
each subcarrier block is given by $N_{0}W$. More specifically,
these diagonal elements correspond to SMC group $j$ and subcarrier
block $n$, and they are associated with either a direct BS-to-UE
SMC  or a BS-to-RN SMC. The additional subscripts
$e_{1}\in\{0,\cdots,\min[N_{B},K\cdot\min(N_{B},N_{U})]\}$ and $e\in\{0,\cdots,\min[N_{B},M\cdot\min(N_{B},N_{R}),K\cdot\min(N_{R},N_{U})]\}$
are used for distinguishing the multiple selected SMCs of the direct
links (i.e. those related to UEs), from the multiple selected SMC-pairs%
\footnote{A single SMC-pair consists of a SMC for the first phase and another
for the second phase. Although these SMCs are generated separately
in each phase, the SMC-pair associated with a common RN has to be
considered as a single entity in the SMC grouping algorithms presented
in Section \ref{sec:SGAlgor}. %
} that may be associated with a particular RN $\mathcal{M}(e)$, respectively.
Note that $\mathcal{M}\left(e\right)$ is a function of $e$, representing
the RN index~(similar to $m$ used before) associated with the SMC-pair
$e$, as further detailed in Section~\ref{sec:SGAlgor}.

\vspace{-5mm}
\subsection{BF design in the second transmission phase\label{sub:BFT2}}

The second transmission phase may be characterized
by the \emph{MIMO interference channel}. A similar methodology is
employed in the second transmission phase, except that now both the
BS and the RNs are transmitters, while a number of UEs are receiving.
In this phase, our aim is 1) to design ZFBF matrices for the BS and
RNs to avoid interference between data streams, 2) and to design
a receive-BF matrix for each UE so that the effective channel matrices
associated with each of its transmitters contain
rows which satisfy the semi-orthogonal condition~(\ref{eq:semi-ortho})
for a given $\alpha$. This means that more data streams may be served
simultaneously, thus improving the attainable system performance.
Since there are multiple \emph{distributed} transmitters/MIMO
channel matrices associated with each UE, the SVD method described
in Section~\ref{sub:BFT1}, which is performed in a centralized fashion,
cannot be readily applied at the transmitter side.
Instead, we aim for minimizing the resultant correlation between the
generated SMCs, thus increasing the number of SMCs which satisfy~(\ref{eq:semi-ortho})
for a given $\alpha$. To accomplish this goal, we begin by introducing
the shorthand of $\underline{\mathbf{H}}_{n,k}^{BU,T_{2}}=\mathbf{R}_{n,k}^{U,T_{2}}\mathbf{H}_{n,k}^{BU}$
and $\underline{\mathbf{H}}_{n,m,k}^{RU,T_{2}}=\mathbf{R}_{n,k}^{U,T_{2}}\mathbf{H}_{n,m,k}^{RU}$
as the effective channel matrices between the BS and UE $k$, and
between RN $m$ and UE $k$, respectively, on subcarrier block $n$
in the second transmission phase, where $\mathbf{R}_{n,k}^{U,T_{2}}$
is the yet-to-be-determined UE $k$'s receive-BF matrix. In light
of the preceding discussions, one of our aims is to design $\mathbf{R}_{n,k}^{U,T_{2}}$
so that the off-diagonal values of the matrices given by $\mathbf{A}_{0}=\underline{\mathbf{H}}_{n,k}^{BU,T_{2}}\left(\underline{\mathbf{H}}_{n,k}^{BU,T_{2}}\right)^{\mathrm{H}}$
and $\mathbf{A}_{m}=\underline{\mathbf{H}}_{n,m,k}^{RU,T_{2}}\left(\underline{\mathbf{H}}_{n,m,k}^{RU,T_{2}}\right)^{\mathrm{H}}\mbox{, }\forall m$
are as small as possible. This design goal may be formalized as 
\begin{eqnarray}
\underset{\mathbf{R}_{n,k}^{U,T_{2}}}{\mbox{min. }} &  & \left|\left|\mathbf{H}_{n,k}^{BU}\left(\mathbf{H}_{n,k}^{BU}\right)^{\mathrm{H}}-\left(\mathbf{R}_{n,k}^{U,T_{2}}\right)^{-1}\boldsymbol{\Lambda}_{0}\left(\mathbf{R}_{n,k}^{U,T_{2}}\right)^{\mathrm{-H}}\right|\right|_{\mathrm{F}}^{2}\nonumber \\
 &  & +\sum_{m=1}^{M}\left|\left|\mathbf{H}_{n,m,k}^{RU}\left(\mathbf{H}_{n,m,k}^{RU}\right)^{\mathrm{H}}\right.\right.\nonumber \\
 &  & \left.\left.-\left(\mathbf{R}_{n,k}^{U,T_{2}}\right)^{-1}\boldsymbol{\Lambda}_{m}\left(\mathbf{R}_{n,k}^{U,T_{2}}\right)^{\mathrm{-H}}\right|\right|_{\mathrm{F}}^{2},\label{eq:diagonalization}
\end{eqnarray}
where $\boldsymbol{\Lambda}_{0}$ and $\boldsymbol{\Lambda}_{m}$
are diagonal matrices containing the diagonal elements of $\mathbf{A}_{0}$
and $\mathbf{A}_{m}$, respectively. Therefore, $\left(\mathbf{R}_{n,k}^{U,T_{2}}\right)^{-1}$
is the \emph{jointly diagonalizing matrix}~\cite{Yeredor2002}, while
$\mathbf{H}_{n,k}^{BU}\left(\mathbf{H}_{n,k}^{BU}\right)^{\mathrm{H}}$
and $\mathbf{H}_{n,m,k}^{RU}\left(\mathbf{H}_{n,m,k}^{RU}\right)^{\mathrm{H}}$,
$\forall m$ are the matrices to be diagonalized. Thus, the algorithm
presented in~\cite{Yeredor2002} for solving~(\ref{eq:diagonalization}) may be invoked at
UE $k$ for obtaining $\mathbf{R}_{n,k}^{U,T_{2}}$, which may be
further fed back to the BS and RNs. Hence, the BS and RNs do not have
to share ${\bf H}_{n,k}^{BU}$ or ${\bf H}_{n,k}^{RU}$ via the wireless
channel and do not have to solve~(\ref{eq:diagonalization}) again.
As a result, we accomplish the goal of creating effective
channel matrices that contain rows aiming to satisfy~(\ref{eq:semi-ortho}).
Additionally, the columns of $\mathbf{R}_{n,k}^{U,T_{2}}$
have been normalized so that the power assigned for each SMC remains
unaffected.

After obtaining the receive-BF matrix, the SMCs of the transmissions
to UE $k$ on subcarrier block $n$ are given by
the non-zero rows of the effective channel matrices $\underline{\mathbf{H}}_{n,k}^{BU,T_{2}}$
and $\underline{\mathbf{H}}_{n,m,k}^{RU,T_{2}}$, $\forall m$. Since
the BS and the RNs act as distributed broadcasters in the second phase,
they are only capable of employing \emph{separate}
ZFBF transmit matrices to ensure that none of them imposes interference
on the SMCs it does not explicitly intend to serve. By employing one
of the grouping algorithms described in Section~\ref{sec:SGAlgor},
the BS schedules $Q_{j}^{T_{2}}\leq\min\left[\min\left(N_{B},N_{R}\right),\sum_{i=1}^{K}L_{i}^{B}+L_{i}^{R}\right]$
SMCs to serve simultaneously in the second phase, where $L_{i}^{B}$
and $L_{i}^{R}$ represent the number of SMCs of UE
$i$ served by the BS and by RNs in this phase, respectively, where
we have $L_{i}^{B}+L_{i}^{R}\le N_{U}$, $L_{i}^{B}\le\min(N_{B},N_{U})$,
and $L_{i}^{R}\le\min(N_{R},N_{U})$. Let
us denote the \emph{refined} effective DL channel matrices, from
the perspectives of the BS and RN $m$, consisting of the $Q_{j}^{T_{2}}$
selected SMCs as $\underline{\mathbf{H}}_{n,j}^{B,T_{2}}$
and $\underline{\mathbf{H}}_{n,j,m}^{R,T_{2}}$,
respectively. Since these are known to each transmitter, they may
employ ZFBF transmit matrices in the second phase, given by the right
inverses $\mathbf{T}_{n,j}^{B,T_{2}}=\left(\underline{\mathbf{H}}_{n,j}^{B,T_{2}}\right)^{\mathrm{H}}\cdot\left[\underline{\mathbf{H}}_{n,j}^{B,T_{2}}\left(\underline{\mathbf{H}}_{n,j}^{B,T_{2}}\right)^{\mathrm{H}}\right]^{-1}$
for the BS, and $\mathbf{T}_{n,j,m}^{R,T_{2}}=\left(\underline{\mathbf{H}}_{n,j,m}^{R,T_{2}}\right)^{\mathrm{H}}\cdot\left[\underline{\mathbf{H}}_{n,j,m}^{R,T_{2}}\left(\underline{\mathbf{H}}_{n,j,m}^{R,T_{2}}\right)^{\mathrm{H}}\right]^{-1}$
for RN $m$. Similar to the first transmission phase,
these ZFBF transmit matrices are normalized by $\mathbf{W}_{n,j}^{BU,T_{2}}$
and $\mathbf{W}_{n,j,m}^{RU,T_{2}}$, respectively, to ensure that
each SMC transmission is initially set to unit power. Upon
obtaining the selected SMCs, we denote the effective
CNRs in the second transmission phase as $G_{n,j,e_{2}}^{BU,T_{2}}=\left|w_{n,j,e_{2}}^{BU,T_{2}}\right|^{2}/\Delta\gamma N_{0}W$
and $G_{n,j,e}^{RU,T_{2}}=\left|w_{n,j,e}^{RU,T_{2}}\right|^{2}/\Delta\gamma N_{0}W$,
where $w_{n,j,e_{2}}^{BU,T_{2}}$ and $w_{n,j,e}^{RU,T_{2}}$ are
the diagonal elements in $\mathbf{W}_{n,j}^{BU,T_{2}}$
and $\mathbf{W}_{n,j,\mathcal{M}(e)}^{RU,T_{2}}$, respectively, and
the subscript $\mathcal{M}(e)$ has been defined in Section~\ref{sub:BFT1}.
To elaborate, for a second-phase BS-to-UE link, $w_{n,j,e_{2}}^{BU,T_{2}}$
corresponds to SMC group $j$ and subcarrier block $n$, while the
subscript $e_{2}\in\{0,\cdots,\min[N_{B},K\cdot\min(N_{B},N_{U})]\}$
is employed for further distinguishing the multiple selected SMCs
associated with UEs from the BS. Similarly, $w_{n,j,e}^{RU,T_{2}}$,
which also corresponds to SMC group $j$ and subcarrier
block $n$, is associated with the second-phase RN-to-UE link between
RN $\mathcal{M}\left(e\right)$ and the particular UE of SMC-pair $e$. 
\vspace{-3mm}
\section{Semi-orthogonal grouping algorithms\label{sec:SGAlgor}}

As described in Section~\ref{sec:SysModel}, the BS has to choose
$Q_{j}^{T_{1}}$ and $Q_{j}^{T_{2}}$ SMCs for the first and second
transmission phases, respectively.
These selected SMCs collectively form the SMC group $j$.
Since the system supports both direct and relaying links, the grouping
algorithms described in~\cite{Yoo2006,UlHassan2009}, which were
designed for MIMO systems dispensing with relays, may not be directly
applied. Instead, we propose a pair of viable grouping algorithms,
namely the exhaustive search-based grouping algorithm~(ESGA), and
the orthogonal component-based grouping algorithm~(OCGA).

Furthermore, for greater flexibility in forming viable SMC groups, additional SMCs may be considered in the
second transmission phase, when tentatively assuming that only a subset of transmitters are
activated. By employing this full list of SMCs, the system can achieve a higher performance.

In both grouping algorithms, each particular SMC must be evaluated before it
may be included into the SMC group to be generated. This evaluation process is completed by
the \emph{SMCCheck}\footnote{More details related to this algorithm may be found in~\cite{Cheung2014}.} algorithm, which
ensures that the SMC to be grouped satisfies the semi-orthogonality criterion of~(\ref{eq:semi-ortho}), while the transmit
and receive dimensions of all nodes and the maximum spatial multiplexing gains of both transmission phases are not exceeded.
\vspace{-3mm}
\subsection{ESGA and OCGA}

\begin{algorithm} \small \SetKw{Or}{or} \SetKw{True}{true} \SetKw{False}{false} \SetKwFunction{SMCCheck}{SMCCheck} \SetKwFunction{ESGA}{ESGA} \SetKwInOut{Input}{inputs}\SetKwInOut{Output}{outputs}
\Input{set of SMC groups associated with subcarrier block $n$ (initialized as empty set), $\mathcal{G}_{n}$ \newline current SMC group (initialized as empty set), $\mathcal{E}_{n,j}$ \newline SMCs associated with subcarrier block $n$, $\mathcal{E}_{n}$ \newline semi-orthogonality parameter $\alpha$} \Output{none} \BlankLine void \ESGA$\left(\mathcal{G}_{n},\mathcal{E}_{n,j},\mathcal{E}_{n},\alpha\right)$ \BlankLine \Begin{ 	\BlankLine 	\ForEach{$e_{c}\in \mathcal{E}_{n} $} 	{ \label{algo:ESGA:enumerate} 		\If{\SMCCheck$\left(e_{c},\mathcal{E}_{n,j},\alpha\right)$ } 		{ \label{algo:ESGA:SMC_check} 			$\mathcal{E}'_{n,j'}\leftarrow \mathcal{E}_{n,j} \cup \{e_{c}\}$\; \label{algo:ESGA:new_group} 			$\mathcal{G}_{n} \leftarrow \mathcal{G}_{n} \cup \{\mathcal{E'}_{n,j'}\}$\;\label{algo:ESGA:append} 			\ESGA$\left(\mathcal{G}_{n},\mathcal{E}'_{n,j'},\mathcal{E}_{n}\setminus e_c,\alpha\right)$\; \label{algo:ESGA:recursion} 		} 	}\label{algo:end_ESGA_enumerate} 	\KwRet\; }
\caption{Exhaustive search-based grouping algorithm~(ESGA)} \label{algo:ESGA} \end{algorithm}

\begin{algorithm} \small \SetKw{Or}{or} \SetKw{True}{true} \SetKw{False}{false} \SetKwFunction{SMCCheck}{SMCCheck} \SetKwFunction{OCGA}{OCGA} \SetKwFunction{NOC}{NOC} \SetKwInOut{Input}{inputs}\SetKwInOut{Output}{outputs}
\Input{set of SMC groups associated with subcarrier block $n$ (initialized as empty set), $\mathcal{G}_{n}$ \newline current SMC group (initialized as empty set), $\mathcal{E}_{n,j}$ \newline SMCs associated with subcarrier block $n$, $\mathcal{E}_{n}$ \newline semi-orthogonality parameter $\alpha$} \Output{none} \BlankLine void \OCGA$\left(\mathcal{G}_{n},\mathcal{E}_{n,j},\mathcal{E}_{n},\alpha\right)$ \BlankLine \Begin{ 	\BlankLine 	complete $\leftarrow$ \True\; 	$\mathcal{E}_{c}\leftarrow \{\}$\; \label{algo:OCGA:candidate_set} 	\BlankLine 	\ForEach{$e_{c}\in \mathcal{E}_{n} $} 	{ \label{algo:OCGA:enumerate} 		\If{\SMCCheck$\left(e_{c},\mathcal{E}_{n,j},\alpha\right)$ } 		{ \label{algo:OCGA:SMC_check} 			\uIf{$\left|\mathcal{E}_{n,j}\right|==0$} 			{\label{algo:empty_check} 				$\mathcal{E}'_{n,j'}\leftarrow \mathcal{E}_{n,j} \cup \{e_{c}\}$\; \label{algo:OCGA:new_set_begin} 				\OCGA$\left(\mathcal{G}_{n},\mathcal{E}'_{n,j'},\mathcal{E}_{n}\setminus e_c,\alpha\right)$\; 				\KwRet\; \label{algo:OCGA:new_set_end} 			} 			\Else 			{ 				$\mathcal{E}_{c}\leftarrow \mathcal{E}_{c} \cup \{e_{c}\}$\; \label{algo:OCGA:add_candidate} 				complete $\leftarrow$ \False\; 			} 		} 	}\label{algo:end_of_creatiing_candidate_group}
	\BlankLine 	\uIf{complete} 	{ 		$\mathcal{G}_{n} \leftarrow \{\mathcal{E}_{n,j}\}$\;\label{algo:OCGA:cannot_find}  	} 	\Else 	{ 		$\mathcal{E}'_{n,j'}\leftarrow \mathcal{E}_{n,j} \cup \underset{e_{c}\in\mathcal{E}_{c}}{\arg\max}$ \NOC$\left(e_{c},\mathcal{E}_{n,j}\right)$\; \label{algo:OCGA:best_candidate} 		\OCGA$\left(\mathcal{G}_{n},\mathcal{E}'_{n,j'},\mathcal{E}_{n}\setminus e_c,\alpha\right)$\; \label{algo:OCGA:recursive} 	}
\BlankLine \KwRet\; }
\caption{Orthogonal component-based grouping algorithm~(OCGA)} \label{algo:OCGA} \end{algorithm}

We present our first grouping method in Algorithm~\ref{algo:ESGA}.
Simply put, the ESGA recursively creates new SMC groups by exhaustively
searching through all the possible combinations of SMCs and including
those that pass the SMC checking algorithm. To elaborate, in
the loop ranging from line~\ref{algo:ESGA:enumerate} to line~\ref{algo:end_ESGA_enumerate},
the algorithm searches through all the possible SMCs associated with
subcarrier block $n$, which are collectively denoted
by $\mathcal{E}_{n}$ and satisfy $e_{c}\in\mathcal{E}_{n}$. The
specific SMCs that satisfy the checks performed in
line~\ref{algo:ESGA:SMC_check} are appended to the current SMC group
in line~\ref{algo:ESGA:new_group}, and the resultant updated SMC
group $\mathcal{E}'_{n,j'}$  is appended to the
set of SMC groups obtained for subcarrier block $n$ in line~\ref{algo:ESGA:append}.
Additionally, $\mathcal{E}'_{n,j'}$ is used recursively in line~\ref{algo:ESGA:recursion}
for filling this group and for forming new groups. The computational
complexity of ESGA is dependent on the number of SMCs which are semi-orthogonal
to each other. The worst-case complexity is obtained when every SMC
satisfies the checks performed in line~\ref{algo:ESGA:SMC_check},
leading to a time-complexity (in terms of the number
of SMC groups generated) upper-bounded (not necessarily tight) by
$\mathcal{O}\left(\sum_{n=1}^{N}\left|\mathcal{E}_{n}\right|^{\theta}\right)$,
where $\theta = \min\left[N_{B},K\cdot\min\left(N_{B},N_{U}\right)+M\cdot\min\left(N_{B},N_{R}\right)\right]+\min\left[\min\left(N_{B},N_{R}\right),\sum_{i=1}^{K}L_{i}^{B}+L_{i}^{R}\right].$

In other words, each subcarrier block may be treated independently.
For each subcarrier block, $\left|\mathcal{E}_{n}\right|$ SMCs must
be checked until the maximum multiplexing gain in both the first and
second phases has been attained.

The second algorithm, OCGA, is presented in Algorithm~\ref{algo:OCGA},
which aims to be a lower complexity alternative to ESGA. The OCGA
commences by creating a SMC candidate set $\mathcal{E}_{c}$,
whose elements satisfy the checks performed in the \emph{SMCCheck} algorithm,
in lines~\ref{algo:OCGA:candidate_set} to~\ref{algo:end_of_creatiing_candidate_group}.
More specifically, if the current SMC group $\mathcal{E}_{n,j}$
is empty, the algorithm can simply create a new SMC group containing
only the candidate SMC that has passed the \emph{SMCCheck} algorithm in lines~\ref{algo:empty_check}
to~\ref{algo:OCGA:new_set_end}. If the SMC group is not empty, the
algorithm adds to it the particular SMC candidate that results in
the highest norm of the orthogonal component~(NOC), via the Gram-Schmidt
procedure~\cite{Yoo2006,UlHassan2009}, in line~\ref{algo:OCGA:best_candidate}.
This process is repeated until the maximum multiplexing gain in both
the first and second phases has been attained. When comparing the
NOCs obtained for the relaying links, the minimum of the NOCs obtained
from the BS-to-RN and RN-to-UE SMCs is used. This is because the information
conveyed on the relaying link is limited by the weaker of the two
transmissions, which is reflected in the effective channel gains quantified
by these norms. If no SMCs satisfy the checks of
line~\ref{algo:OCGA:SMC_check}, the current SMC group is complete,
and it is appended to the current set of SMC groups in line~\ref{algo:OCGA:cannot_find}.
Since new groups are only created when the current SMC group is empty,
this algorithm results in much fewer groups than ESGA. The algorithmic
time-complexity is given by $\mathcal{O}\left(\sum_{n=1}^{N}\left|\mathcal{E}_{n}\right|\right)$
as a single group is created for each initially-selected SMC.

Both grouping algorithms may be initialized with
an empty SMC group, $\mathcal{E}_{n,j}\leftarrow\left\{ \right\} $,
and an empty set of SMC groups, $\mathcal{G}_{n}\leftarrow\left\{ \right\} $
, so that they recursively create and fill SMC groups according to
their criteria. Additionally, a final step is performed
to remove the specific groups, which result in effective channel gains
that are less than or equal to that of another group, while having
the same transmitters. Therefore, this final step does not reduce
the attainable system performance, but reduces the number of possible groups,
thus alleviating computational complexity.
\vspace{-3mm}
\section{Numerical results and discussions\label{sec:ResDis}}

\begin{table}[t]
\centering{}\protect\caption{Simulation parameters used to obtain all results in Section~\ref{sec:ResDis}
unless otherwise specified.}

\label{tab:param} \setlength{\extrarowheight}{1.5pt} %
\begin{tabular}{|l|r|}
\hline 
\textbf{Simulation parameter}  & \textbf{Value}\tabularnewline
\hline 
\hline 
Subcarrier block bandwidth, $W$ {[}Hertz{]}  & $180$k\tabularnewline
\hline 
Antenna configuration, $\left(N_{B},N_{R},N_{U}\right)$  & $\left(4,4,2\right)$\tabularnewline
\hline 
Cell radius, {[}km{]}  & $\{0.75,1.75\}$\tabularnewline
\hline 
Ratio of BS-to-RN distance to the cell radius & $0.5$\tabularnewline
\hline 
SNR gap of wireless transceivers, $\Delta\gamma$ {[}dB{]}  & 0\tabularnewline
\hline 
Noise power spectral density, $N_{0}$ {[}dBm/Hz{]}  & \textminus 174\tabularnewline
\hline 
Number of channel samples  & $10^{4}$\tabularnewline
\hline 
\end{tabular}

\vspace{-1mm}
\end{table}

This section presents the numerical results obtained, when employing
the grouping algorithms described in Section~\ref{sec:SGAlgor} to the MIMO-OFDMA multi-relay
cellular network considered. The pertinent simulation parameters are
given in Table~\ref{tab:param}. Additionally, the path-loss effect
is characterized relying on the method and parameters
of~\cite{3GPP_PL}, where the BS-to-UE and RN-to-UE
links are assumed to be non-line-of-sight~(NLOS) links, since these
links are typically blocked by buildings and other large obstructing
objects, while the BS-to-RN links are realistically
assumed to be line-of-sight~(LOS) links, as the RNs may be strategically
deployed on tall buildings to create strong wireless backhaul links.
Furthermore, independently and randomly generated set of UE locations
as well as fading channel realizations were used for each channel
sample.
\vspace{-5mm}
\subsection{On the optimality and the relative complexity of ESGA and OCGA for
various $\alpha$ values}

\begin{figure}
\centering
\includegraphics[scale=0.9]{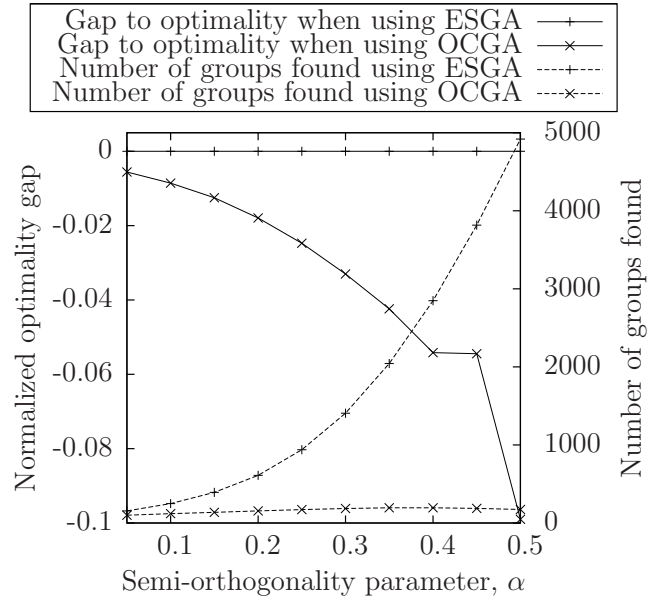} 
\caption{The optimality gap and total number of SMC groups
found when employing the ESGA and OCGA, and using the
parameters in Table~\ref{tab:param} with $N=6$, $K=2$, $M=2$,
$P_{max}^{B}=20$dBm, $P_{max}^{R}=10$dBm and a cell radius of $0.75$km.}
\label{fig:complexity}
\vspace{-5mm}
\end{figure}

Firstly, the behavior of the ESGA
and OCGA as a function of $\alpha$ is examined. Note that in Fig.~\ref{fig:complexity}
the optimal system capacity is attained, when employing the maximization algorithm of~\cite{Cheung2014}, since the ESGA is capable of
enumerating all possible SMC groupings satisfying~(\ref{eq:semi-ortho})
for the corresponding $\alpha$. The 'normalized
optimality gap' is then defined as $\left(\beta/\beta^{*}\right)-1$,
where $\beta^{*}$ is the optimal capacity obtained from employing the ESGA
algorithm, and $\beta$ is the capacity obtained from (in this case) the OCGA
algorithm. We can see from Fig.~\ref{fig:complexity},
that the normalized optimality gap of OCGA relative to ESGA is about
$-0.005\sim-0.1$ for the $\alpha$ values considered.
However, the number of groups found using ESGA is exponentially increasing
with $\alpha$. By contrast, for OCGA, this number
is always significantly lower and gradually becomes less than $200$,
when $\alpha$ increases to $0.5$. In fact, the number of groups
found by OCGA is reduced to about $3.5\%$ of that found by ESGA
at $\alpha=0.5$. This demonstrates the viability of employing OCGA as a reduced-complexity near-optimum alternative
to ESGA.
\vspace{-5mm}
\subsection{The variation in achievable capacity for different values of $P_{max}^{B}$
and $P_{max}^{R}$}

\begin{figure}
\centering
\subfloat[Average achievable capacity for varying $P_{max}^{B}$.]{\includegraphics[scale=0.82]{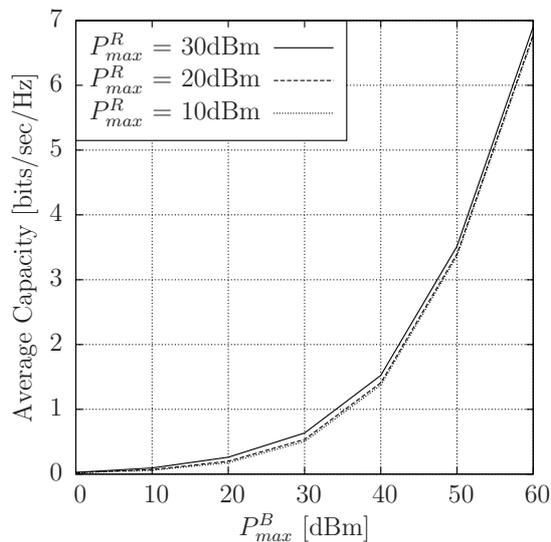}}

\subfloat[Average achievable capacity for varying $P_{max}^{R}$.]{\includegraphics[scale=0.82]{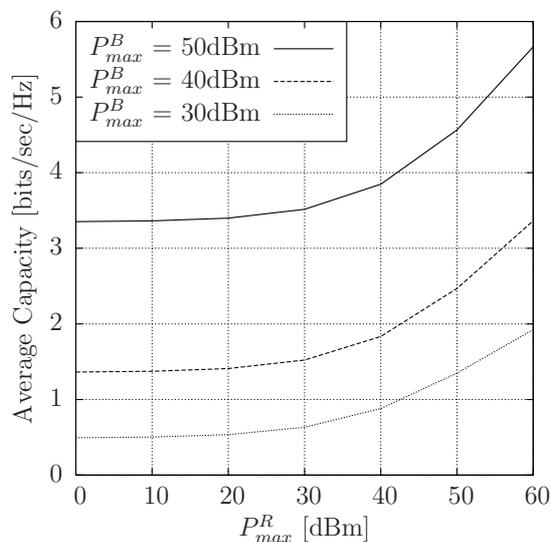}}

\caption{The average achievable capacity of the OCGA with random group selection and equal power
allocation. The parameters in Table~\ref{tab:param} with $N=6$, $K=10$, $M=2$,
$\alpha=0.1$ and a cell radius of $1.75$km are used.}
\label{fig:powers}
\vspace{-7mm}
\end{figure}

As shown in Fig.~\ref{fig:powers}(a), the achievable capacity is monotonically
increasing with $P_{max}^{B}$ and $P_{max}^{R}$, when employing
the OCGA in conjunction with random group selection and equal power allocation. This is
expected, since the capacity is a monotonically increasing function of the assigned
power. Furthermore, it is clear that the effect of increasing $P_{max}^{B}$
on the capacity is significantly more pronounced, than that of applying
the same increase to $P_{max}^{R}$. The intuitive reasoning behind
this is that the power available at the BS has a more pronounced effect
on the system's performance, since the direct links and, more importantly,
the BS-to-RN links rely on the BS. Therefore, increasing $P_{max}^{R}$
is futile, if the BS-to-RN links are not allocated sufficient power
to support the RN-to-UE links.
\vspace{-3mm}
\section{Conclusions\label{sec:Conc}}

In this paper, a novel transmission protocol based on joint
transmit-BF and receive-BF was developed for the multi-relay MIMO-OFDMA
cellular network considered. By employing this protocol,
the MIMO channel matrices were mathematically decomposed into several SMCs,
which may be grouped for transmission to attain a high multiplexing gain. Therefore,
we proposed both an exhaustive grouping algorithm and a lower-complexity alternative.
These algorithms were evaluated based on their performance versus complexity trade-off.
Furthermore, our additional results demonstrated the different effects that the available power
at the BS and the RNs have on the system's capacity.
\vspace{-3mm}
\bibliographystyle{IEEEtran}
\bibliography{references}
\end{document}